\documentclass[useAMS,fleqn,usenatbib]{mnras}
\usepackage[utf8]{inputenc}
\usepackage{amssymb,amsmath}
\usepackage{graphicx}
\usepackage[dvipsnames]{xcolor}
\hypersetup{colorlinks=true,allcolors=teal}
\usepackage[flushleft]{threeparttable}
\usepackage{footmisc}

\newcommand{\be}{\begin{equation}}
\newcommand{\ee}{\end{equation}}
\def\bear#1\ear{\begin{align}#1\end{align}}

\newcommand{\fig}[1]{Fig.~\ref{#1}}

\usepackage{newtxtext,newtxmath}

\usepackage[T1]{fontenc}

\DeclareRobustCommand{\VAN}[3]{#2}
\let\VANthebibliography\thebibliography
\def\thebibliography{\DeclareRobustCommand{\VAN}[3]{##3}\VANthebibliography}

\usepackage{graphicx}	
\usepackage{amsmath}	
\usepackage{subfig}

\title[GPR trained SCRIPT]{A fast method of reionization parameter space exploration using GPR trained SCRIPT}

\author[Maity, Paranjape \& Choudhury]{
Barun Maity$^{1}$\thanks{E-mail: bmaity@ncra.tifr.res.in}, Aseem Paranjape$^{2}$, and Tirthankar Roy Choudhury$^{1}$\\
$^{1}$National Centre for Radio Astrophysics (NCRA), TIFR, Pune University Campus, Post Bag 3, Pune 411 007, India\\
$^{2}$Inter University Centre for Astronomy and Astrophysics (IUCAA), Pune University Campus, Post Bag 4, Pune 411 007, India}
\date{Accepted XXX. Received YYY; in original form ZZZ}

\pubyear{2023}

\begin{document}
\label{firstpage}
\pagerange{\pageref{firstpage}--\pageref{lastpage}}
\maketitle

\begin{abstract}
Efficient exploration of parameter spaces is crucial to extract physical information about the Epoch of Reionization from various observational probes. To this end, we propose a fast technique based on Gaussian Process Regression (GPR) training applied to a semi-numerical photon-conserving reionization model, \texttt{SCRIPT}. Our approach takes advantage of the numerical convergence properties of \texttt{SCRIPT} and constructs a training set based on low-cost, coarse-resolution simulations. A likelihood
emulator is then trained using this set to produce results in approximately two orders of magnitude less computational time than a full MCMC run, while still generating reasonable 68\% and 95\% confidence contours. Furthermore, we conduct a forecasting study using simulated data to demonstrate the applicability of this technique. This method is particularly useful when full MCMC analysis is not feasible due to expensive likelihood computations.
\end{abstract}




\section{Introduction}

Understanding the reionization of neutral hydrogen (HI) in detail is one of the major challenges in modern cosmology. This epoch marks the last major phase transition of the universe when it makes the transition from a cold and neutral state to hot and ionized state. The main drivers of this process are expected to be the first luminous sources in the universe, the nature of which is yet to be completely understood \citep[for details, see][]{2001PhR...349..125B,2009CSci...97..841C,2018PhR...780....1D,2022arXiv220802260G,2022GReGr..54..102C}. Additionally, the exact timeline of the reionization epoch is still not fully understood, but we expect to obtain a clear answer with the help of upcoming observational facilities.

It is a challenging task to decipher the reionization phenomena due to the involvement of complex astrophysical processes. However, there is a wide variety of available  and planned observational facilities spanning multiple wavelength bands, which will be crucial to glean the physical information during the reionization. These probes include the Cosmic Microwave Background (CMB) \citep{2020A&A...641A...6P},  opacity fluctuations of the Lyman-$\alpha$ (Ly$\alpha$) absorption spectra of quasars at $z \lesssim 6$ \citep[e.g.][]{2015MNRAS.447.3402B,2018MNRAS.479.1055B,2018ApJ...864...53E,2019ApJ...881...23E,2020ApJ...904...26Y,2021arXiv210803699B} and the properties of the Ly$\alpha$ emitters \citep[e.g.][]{2014ApJ...797...16K,2017ApJ...844...85O,2017ApJ...842L..22Z,2018PASJ...70S..16K,2018ApJ...867...46I,2018PASJ...70S..13O, 2019ApJ...886...90H, 2021ApJ...919..120M, 2022ApJ...927...36W}. Some of the other indirect observational probes include high redshift Ly$\alpha$ emitters clustering measurements \citep[e.g.][]{2014ApJ...788...87F,2022ApJ...940..102P}, the Lyman break galaxies \citep[e.g.][]{2018ApJ...856....2M,2020ApJ...892..109N}, Ly$\alpha$ damping wings in high-redshift quasar spectra \citep[e.g.][]{2018Natur.553..473B,2022MNRAS.512.5390G}, and the measurements of the effective optical depth of the Ly$\alpha$ and Ly$\beta$ forests \citep[e.g.][]{2015MNRAS.447..499M, 2021ApJ...923..223Z}. These data provide crucial information about the state of the intergalactic medium (IGM). i.e., the ionization history of the universe. On the other hand, the imprint of the reionization sources can be studied using the ultra-violet luminosity function (UVLF) of high-redshift galaxies \citep[e.g.][]{2015ApJ...810...71F,2015ApJ...800...18A,2018MNRAS.479.5184A,2015ApJ...803...34B,2017ApJ...843..129B,2021AJ....162...47B}. The thermal evolution of the IGM has also been proven as an important probe of reionization. Temperature estimates of low density IGM, measured using Ly$\alpha$ absorption spectra \citep{2019ApJ...872...13W,2020MNRAS.494.5091G} have the potential to probe the late phases of the reionization \citep{2022MNRAS.515..617M}.  Lastly, the redshifted 21 cm signal from the spin flip transition of the neutral hydrogen atom will open up a treasure trove of information about this high redshift epoch \citep[e.g.][]{1972CoASP...4..173S,1979MNRAS.188..791H,1990MNRAS.247..510S, 2001JApA...22..293B, 2012RPPh...75h6901P}. There are attempts to measure the global 21~cm signal which carries the global signature of thermal and ionization histories in the universe \citep[EDGES;][]{2018Natur.555...67B}, \citep[SARAS3;][]{2022NatAs.tmp...47S}. However, it misses the information coming from  the fluctuations in ionization field. In complementary, current radio interferometers including Low Frequency Array \citep[LOFAR;][]{2013A&A...556A...2V}, Murchison Widefield Array \citep[MWA;][]{2013PASA...30....7T}, Giant Metrewave Radio Telescope \citep[GMRT;][]{1991ASPC...19..376S, 2017CSci..113..707G} and (partially deployed) Hydrogen Epoch of Reionization Array \citep[HERA;][]{2017PASP..129d5001D}  which target to measure the fluctuations in neutral hydrogen field. These have already placed interesting upper limits on the amplitude of the 21 cm fluctuations \citep[LOFAR;][]{2017ApJ...838...65P,2019MNRAS.488.4271G,2020MNRAS.493.1662M}, \citep[MWA;][]{2020MNRAS.493.4711T,2023MNRAS.521.5120K}, \citep[GMRT;][]{2013MNRAS.433..639P} and \citep[HERA;][]{2022ApJ...925..221A}. In the near future, telescopes like the Square Kilometer Array \citep[SKA-Low;][]{2015aska.confE...1K} and the fully deployed HERA \citep{2017PASP..129d5001D} will target the direct mapping the signal. Also, there exist experiments like Owens Valley Long Wavelength Array \citep[OVRO-LWA][]{2019AJ....158...84E}  and New Extension in Nançay Upgrading LOFAR \citep[NenuFAR][]{2021sf2a.conf..211M} aiming for the detection of 21~cm signal from higher redshift epoch i.e. Cosmic Dawn.

In parallel, considerable progress has been made in developing efficient and realistic models of the HI field during reionization, which are capable of extracting physical information using the available data. On one hand, there are full radiative hydrodyamic simulations that incorporate accurate physics but are computationally expensive and not ideal for parameter space exploration \citep[e.g.][]{2006MNRAS.372..679M,2006MNRAS.369.1625I,2007ApJ...671....1T,2015MNRAS.447.1806G,2016MNRAS.463.1462O,2019MNRAS.483.1029K,2020MNRAS.496.4087O,2022MNRAS.511.4005K, 2022MNRAS.512.4909G,2022arXiv220713098P}. Recently, some of these simulations have been utilized to build up emulators suitable for parameter inferences studies e.g. \texttt{CRADLE} \citep{2019MNRAS.490.1055C}, \texttt{GRIZZLY} emulator \citep{2020MNRAS.493.4728G} and \texttt{PINION}   
 \citep{2023MNRAS.521..902K}.  On the other hand, semi-numerical/analytic models such as \texttt{21cmFAST} \citep{2007ApJ...669..663M,2011MNRAS.411..955M} and \texttt{SIMFAST21} \citep{2010MNRAS.406.2421S},  \texttt{DRAGONS} \citep{2016MNRAS.462..250M,2016MNRAS.462..804G}, \texttt{SCORCH III} \citep{2020ApJ...905..132C},  \texttt{ASTRAEUS} \citep{2021MNRAS.503.3698H} strike a balance between efficiency and accuracy and can be exploited for exploring the unknown parameters. Most of these semi-numerical models are based on the excursion set approach which makes them extremely efficient computationally. However, the excursion set models are associated with the issue of photon number non conservation \citep{2007ApJ...654...12Z, 2011MNRAS.414..727Z,2016MNRAS.460.1801P}, which can also affect the convergence of the large scale 21 cm power spectra at different resolutions. Recently, a possible solution has been provided by an explicitly photon conserving model, named \texttt{SCRIPT} \citep{2018MNRAS.481.3821C, 2022MNRAS.511.2239M}. There also exist detailed photon conserving algorithms incorporating the directionality of photon propagation, such as \texttt{ARTIST} \citep{2019MNRAS.489.5594M}, \texttt{BEoRN} \citep{2023arXiv230515466S}. All these fast models, coupled with Bayesian statistical techniques like Markov Chain Monte Carlo (MCMC), have been proven to be successful in exploring the space of reionization parameters by comparing with the available observational data \citep{2015MNRAS.449.4246G,2017MNRAS.472.2651G,2018MNRAS.477.3217G,2019MNRAS.484..933P,2021MNRAS.506.2390Q,2022MNRAS.515..617M}. Exploiting the recent upper limits on the 21~cm power spectrum measurements from different telescopes, these semi-numerical models have been capable to rule out somewhat extreme models of heating and ionization \citep{2015ApJ...809...62P,2016MNRAS.455.4295G,2020MNRAS.493.4728G,2020MNRAS.498.4178M,2021MNRAS.500.5322G,2021MNRAS.501....1G,2021MNRAS.503.4551G,2022ApJ...924...51A}

However, even these efficient semi-numerical models can become computationally expensive when required to run at relatively high dynamic range. For instance, with the availability of 21 cm data from telescopes like the SKA, high-resolution simulations are necessary to access large Fourier $k$-modes, while the simulation volume must be large enough to meet survey specifications. Since MCMC-based parameter space explorations require running the simulation a large number of times, especially when the number of free parameters is high, it is not practically feasible to conduct such explorations using high dynamic range simulations. To overcome this challenge, several efforts have been made to expedite parameter space exploration through relatively low number of simulation runs. 

Perhaps the simplest approach to speeding up parameter space exploration with less computational cost is the one based on the Fisher matrix. This method requires only a few ($\sim 10$) simulations around a set of fiducial parameters and is useful for understanding parameter degeneracies \citep{2016MNRAS.458.2710E, 2020MNRAS.498.1480S,2022MNRAS.513.1719G,2022arXiv221209797M}, but it cannot produce confidence contours as accurately as a full MCMC analysis, especially towards the low probability tails of the parameter distributions. Another widely used approach is to use the so-called emulators, where simulations are run for a pre-determined set of parameters, and a machine learning algorithm is trained to interpolate the results for other parameter values. Once trained, the emulator can predict the appropriate simulation output at any desired location in the parameter space with negligible computational cost, making MCMC analysis highly efficient. Consequently, emulators have been widely used to constrain reionization parameters through MCMC analysis \citep[e.g.][]{2017ApJ...848...23K,2018MNRAS.475.1213S,2020MNRAS.493.4728G,2020MNRAS.498.4178M,2022JCAP...04..045T,2022arXiv220108205S}. There also exists studies which aim to directly infer reionization parameters utilizing machine learning technique on 21~cm probe \citep[e.g.][]{2017MNRAS.468.3869S, 2019MNRAS.484..282G, 2019MNRAS.490..371D}

One of the main challenges in building these emulators is to prepare an appropriate training set. In situations where the parameters are highly unconstrained, as is often the case with reionization models, one needs to sample them over a wide prior range. Given that the number of training samples would be limited, there is always a risk of under-sampling the region around the best-fit model, which can affect the subsequent likelihood analysis. Hence, for any emulator developed, it is important to demonstrate that they indeed produce the correct confidence contours for a given data vector, at least for some controlled situations where the full MCMC can be completed. \citet{2017ApJ...848...23K} used an adaptive method to create the training set, starting with a sparse and broad parameter range and then iteratively converging near the region of high probability by running an MCMC at every step. The model is trained using the Gaussian Process Regression (GPR) method.  For a three-parameter reionization model, the constraints obtained using the emulator-based MCMC is quite similar to those from the full one, with a training set of $\sim 10^4$ samples. Such iterative method of converging on regions of high probability for accelerated parameter inference using GPR and neural networks has also been explored in the context of weak lensing and galaxy clustering by \citet{2023MNRAS.518.4818B}. On the other hand, \citet{2018MNRAS.475.1213S} used artificial neural networks to build the emulator based on 21~cm power spectra and showed that their emulator can match the full MCMC for the three-parameter reionization model with a training set of only $\sim 100$ samples. However, the priors chosen for creating their training set were relatively narrower. Also, the observables used in our study are different which we discuss in section \ref{sec:obs}.

Building upon earlier attempts, we aim to construct a GPR-based emulator for our semi-numerical reionization model \texttt{SCRIPT}. Our approach differs from the earlier ones in two main ways. First, we take advantage of the numerical convergence of our model with respect to resolution and construct the training set by sampling the posterior distributions obtained from a full MCMC run using low-cost coarse resolution simulations. Second, we emulate the value of the likelihood (or equivalently, the $\chi^2$), instead of the observables, making the training more efficient if the likelihood is a smooth function of the parameters. This avoids having to accurately emulate any features the observables may have, by exploiting the fact that data errors can potentially wash out such features. 
We validate our emulator by comparing the resulting parameter constraints with a full MCMC run using our five-parameter reionization model introduced in \citet{2022MNRAS.515..617M}. To obtain the parameter constraints, we compare our model with existing observations related to reionization, such as the CMB optical depth, galaxy luminosity function at $z \gtrsim 6$, constraints on neutral fraction at $z \sim 6$, and IGM temperature measurements at $z \sim 5.5$. Although the original motivation for developing the emulator was the upcoming 21 cm observations, we do not use them in this work and will postpone their inclusion to a future project.

This paper is organized as follows: In Section \ref{sec:conv_mcmc}, we discuss the conventional MCMC analysis procedure. Specifically, we describe the theoretical framework and provide a brief overview of the \texttt{SCRIPT} model parameters in Section \ref{sec:theory}. In Section \ref{sec:obs}, we introduce the various observational constraints, followed by defining the likelihood in Section \ref{sec:analysis_mcmc}. Then, in Section \ref{sec:build_interp}, we describe the procedure for building the likelihood emulator.
Next, we present a comparison between the original full MCMC and the MCMC with trained interpolator using presently available data in Section \ref{sec:real_dat}. Following this, we perform a forecast study using futuristic data sets along with a discussion of the advantages and constraints of our method in Section \ref{sec:forecast}. Finally, we summarize our main results in Section \ref{sec:conc}. In this paper, the assumed cosmological parameters are $\Omega_M$ = 0.308, $\Omega_{\Lambda}$ = 0.691 $\Omega_b$ = 0.0482, $h$ = 0.678, $\sigma_8$ = 0.829 and $n_s$ = 0.961 \citep{2016A&A...594A..13P}.

\section{Conventional MCMC analysis}
\label{sec:conv_mcmc}

Let us first briefly discuss the reionization model and the steps to constrain the free parameters by comparing with observations using the conventional MCMC (i.e., without any emulators). The discussion in this section closely follows that in \citet{2022MNRAS.515..617M} and is included for completeness.

\begin{figure*}
    \centering
    \includegraphics[width=\textwidth]{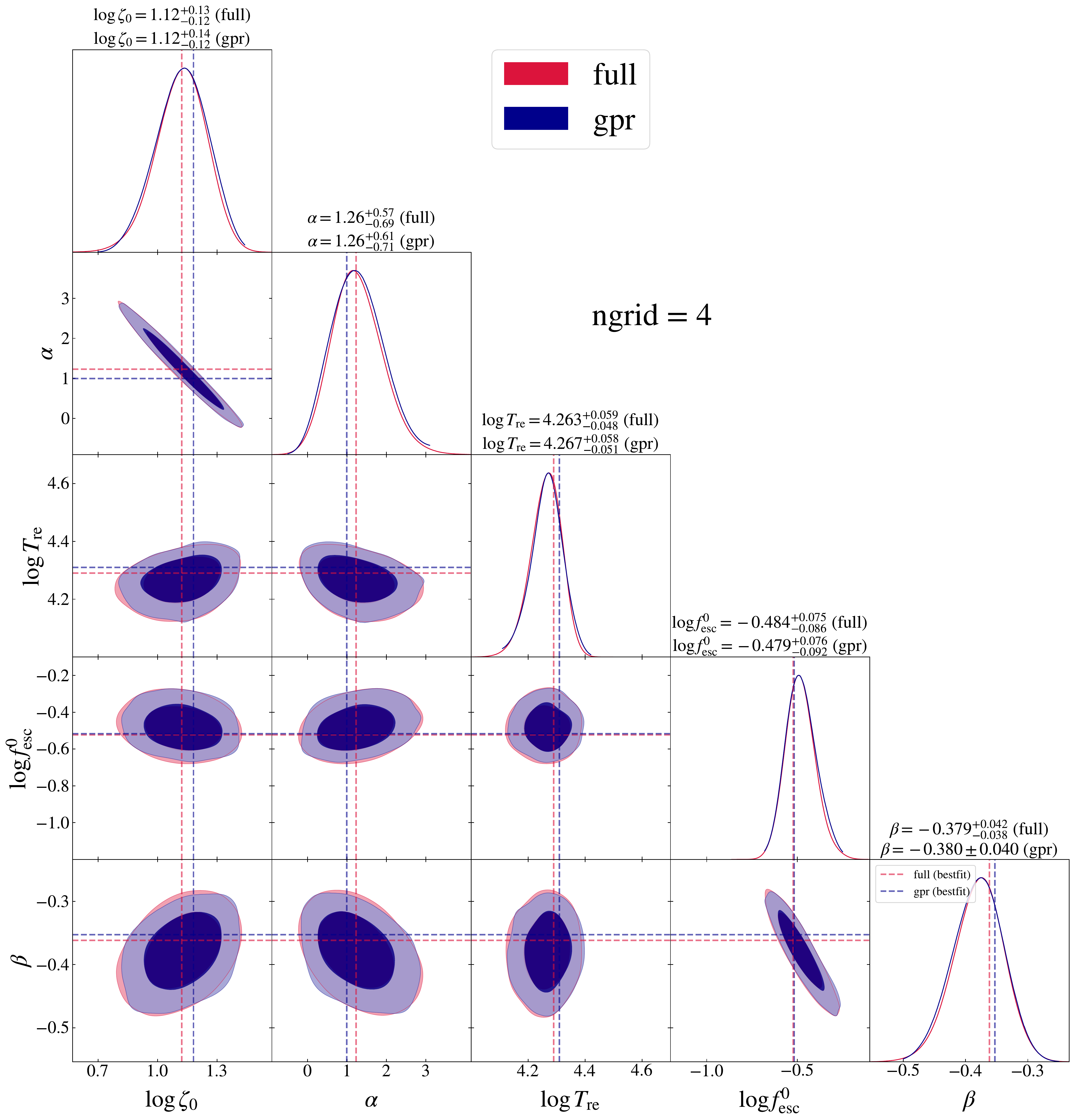}
    \caption{Comparison between full MCMC runs (ngrid=4) and GPR-trained MCMC using a subset of the full run. 
    The diagonal panels show the 1D posterior probability distribution while  off diagonal panels show the joint 2D posteriors. The contours represent 68\% and 95\% confidence intervals. [The dashed lines denote the best-fit values for different cases. The quoted values on each parameter are showing the mean along with $1\sigma$ uncertainties]}
    \label{fig:mcmc_real_dat}
\end{figure*}

\begin{figure*}
    \centering
    \includegraphics[width=\textwidth]{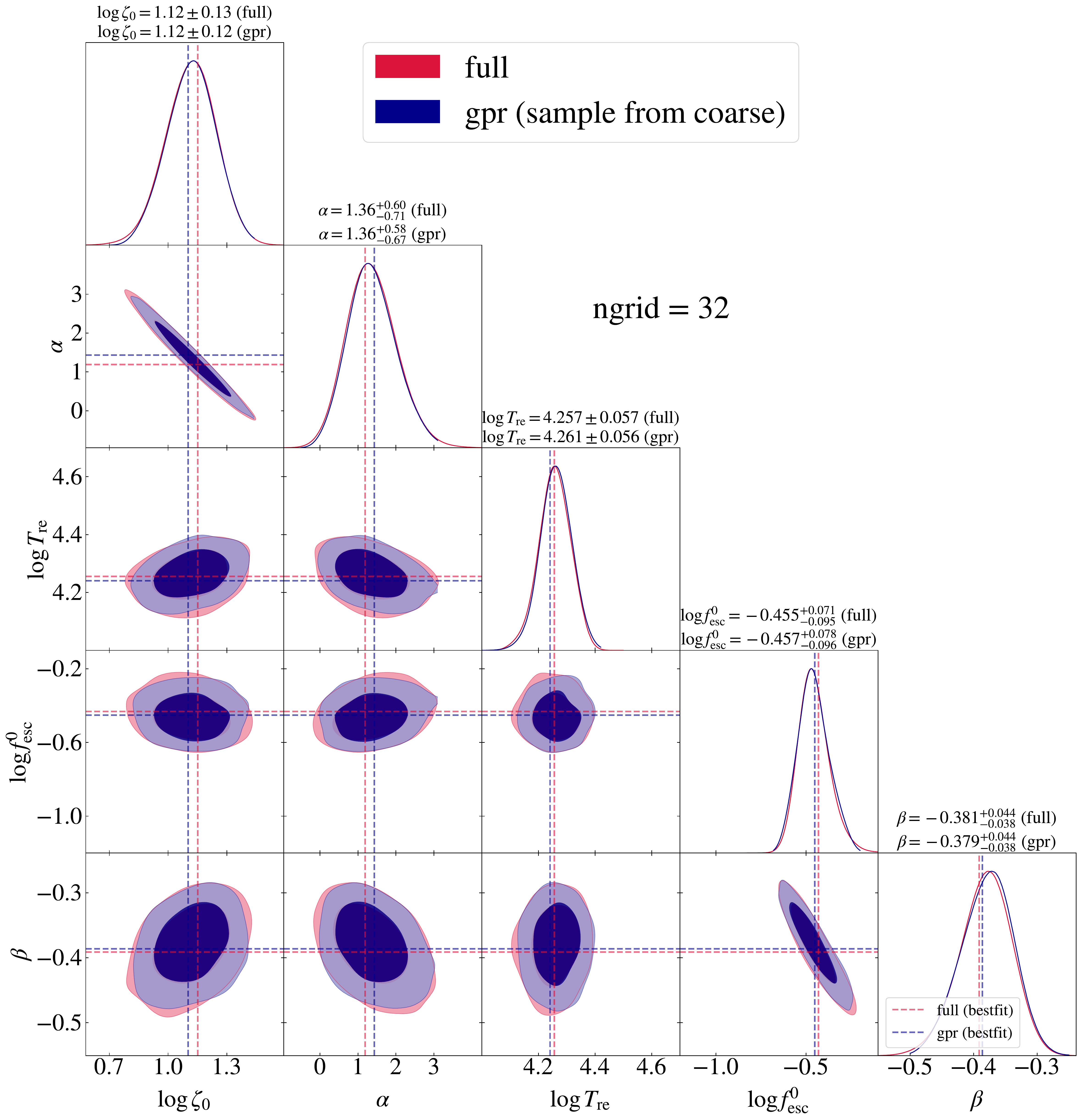}
    \caption{Comparison between full MCMC runs (ngrid=32) and GPR-trained MCMC using a standard smaller set of sample got from earlier coarse resolution analysis. 
    The data are same as used in the coarse resolution run, i.e in \fig{fig:mcmc_real_dat},
    and the formatting is also identical to that Figure.
    }
    \label{fig:mcmc_coarse_fine_jump}
\end{figure*}

\subsection{The reionization model and free parameters}
\label{sec:theory}

For this study, we use semi-numerical photon conserving model \texttt{SCRIPT} of reionization. The models used in \texttt{SCRIPT} are discussed in earlier studies \citep{ 2022MNRAS.511.2239M, 2022MNRAS.515..617M}. The central structure of the model is built upon a semi-numerical photon conserving framework \texttt{SCRIPT} \citep{2018MNRAS.481.3821C}, which provides the ionization state of the universe in a cosmologically representative simulation volume. The main feature of this code is that it conserves the ionizing photons explicitly and hence produces numerically convergent power spectra of ionization fluctuations with respect to the resolution of the ionization maps. Two basic inputs, namely, the density field and the field of collapsed haloes capable of producing ionizing radiation, need to be supplied  to generate the ionization field. As we will be dealing with large scale features of the intergalactic medium (IGM), it is sufficient use the second order Lagrangian perturbation theory approximation \citep[2LPT,][]{1998MNRAS.299.1097S} to generate the density field instead of a full $N$-body simulation. In practice, we use the implementation by \citet{2011MNRAS.415.2101H}.\footnote{\url{https://www-n.oca.eu/ohahn/MUSIC/}} A sub-grid prescription based on the conditional ellipsoidal mass function \citep{2002MNRAS.329...61S} is utilized to compute the halo field. In this work, we use a simulation box size of $256h^{-1}\mathrm{cMpc}$ which is adequate for the observables we calculate for our purpose. From recent studies, it has been shown that the box size $\ge250h^{-1}\mathrm{cMpc}$ is sufficient to provide convergent power spectra considering the observational noise uncertainties \citep{2014MNRAS.439..725I,2020MNRAS.495.2354K}. Our model requires the comoving simulation boxes from $z = 20$ to $z = 5$ with an interval of $\Delta z=0.1$ to compute the full reionization history.

We use the photon conserving algorithm to create the reionization topology in the simulation box. The ionization field requires the ionization efficiency parameter $\zeta(M_h, z)$, which estimates the available ionizing photons per hydrogen atom and can depend on halo mass $M_h$ and redshift $z$. We include inhomogeneous recombinations by tuning the ionization criteria to compensate for excess neutral atoms, and incorporate small-scale fluctuations through a globally averaged clumping factor $C_{\mathrm{HII}}$. We fix the value $C_{\mathrm{HII}} = 3$ in this work motivated from earlier simulation studies \citep[e.g.][]{2020ApJ...898..149D}.

To complement the ionization history, we solve for the thermal history of each grid cell in the box. The code automatically accounts for the effect of spatially inhomogeneous reionization on temperature evolution, assuming that a region's temperature increments by a value $T_{\mathrm{re}}$ as it is ionized for the first time. The parameter $T_{\mathrm{re}}$ is known as the reionization temperature \citep{1997MNRAS.292...27H,2009ApJ...701...94F,2018MNRAS.477.5501K,2022MNRAS.511.2239M}.

 Our method also includes radiative feedback suppressing the production of ionizing photons in haloes where the gas is heated up. In \citet{2022MNRAS.511.2239M}, a variety of methods have been introduced to incorporate radiative feedback effects, we choose to work with the `step feedback' model here. In this case, the gas fraction retained inside the radiative feedback affected halo is assumed to be zero for a halo mass smaller than $M_{\mathrm{min}} = \mathrm{Max} \left[M_{\mathrm{cool}}, M_{J}\right]$,
and unity otherwise, where $M_{\mathrm{cool}}$ is the minimum threshold mass for atomic cooling and  $M_J$ is the Jeans mass at virial overdensity. This is an efficient and simplistic way to incorporate radiative feedback during reionization. $M_{\mathrm{min}}$ is not only redshift dependent but also spatially varying when the feedback effect dominates (in ionized region). This is due to the fact that the $M_J$ is temperature dependent and the different regions have different temperatures following the ionization topology. However, in neutral region (unaffected by feedback), $M_{\mathrm{min}}$ is driven by the atomic cooling mass threshold (i.e. $M_{\mathrm{cool}}$) which has values of $\sim 3\times 10^7 M_{\odot}$ at $z = 20$ and $\sim 2\times 10^8 M_{\odot}$ at $z = 5$. For feedback affected regime, $M_{\mathrm{min}}$ is typically $> 10^9 M_{\odot}$. There exists more sophisticated and realistic model where the feedback effect is gradual assuming a mass dependent depletion of gas fraction instead of a step like cut-off \citep[for details see, ][]{2022MNRAS.511.2239M}, however, these implementations are computationally less efficient compared to the `step feedback' model. The main aim of this work is to demonstrate the suitability of a GPR-based emulator on constraining the model parameters, and the method presented here should be valid for any feedback prescription.

The free parameters of our model are as follows:

\begin{itemize}

\item The ionization efficiency $\zeta$ is assumed to be independent of $M_h$ and its redshift-dependence, motivated by the earlier studies \citep{2015ApJ...813...54T,2016MNRAS.460..417S,2020MNRAS.495.3065D,2021MNRAS.501L...7C,2022MNRAS.511.2239M}, is assumed to be a simple power-law
\be
\label{eq:zeta_eq}
    \zeta(z) = \zeta_0\left(\frac{10}{1+z}\right)^\alpha,
\ee
where $\zeta_0$ is the ionization efficiency at $z = 9$ and $\alpha$ is the slope. We use $\log \zeta_0$ and $\alpha$ as free parameters in our models. 

\item The reionization temperature $T_{\mathrm{re}}$ is critical in the modelling of the IGM temperature evolution, recombination and radiative feedback. We take $\log T_{\mathrm{re}}$ as a free parameter.

\item To compute the UVLF, we need the fraction of photons escaping the halos, $f_{\mathrm{esc}}(M_h, z)$. We utilize UVLF data only at redshifts $z = 6$ and $7$, so we assume that $f_{\mathrm{esc}}$ is solely $M_h$-dependent \citep{2015MNRAS.451.2544P,2016ApJ...833...84X,2020MNRAS.498.2001M} and neglect its redshift evolution from $z = 7$ to $6$. We assume the form 
\be\label{eq:esc_eq}
     f_{\mathrm{esc}} = f_{\mathrm{esc}}^0\left(\frac{M_h}{10^9M_{\odot}}\right)^\beta,
\ee
where $f_{\mathrm{esc}}^0$ is the escape fraction for a halo of mass $10^9 M_{\odot}$ and $\beta$ is the power law index. The ionization efficiency $\zeta \propto f_*~f_{\mathrm{esc}}$ is assumed to be mass-independent, so the mass dependence of star formation efficiency ($f_*$) exactly compensates for that of $f_{\mathrm{esc}}$ and is automatically assumed to be $M_h^{-\beta}$. This choice is consistent with parameter estimates from earlier studies \citep{2019MNRAS.484..933P,2021MNRAS.506.2390Q}.

\end{itemize}

So, we have five free parameters $\mathbf{a} = \{\log \zeta_0, \alpha, \log T_{\mathrm{re}}, \log f_{\mathrm{esc}}^0, \beta\}$ for the various analyses we pursue in this work.

\subsection{Observational Data}
\label{sec:obs}

The observational constraints used in this work are identical to those in \citet{2022MNRAS.515..617M}, as summarized below:

\begin{enumerate}

\item We use the CMB scattering optical depth ($\tau_e$) of $\tau_e = 0.054 \pm 0.007$ from the latest Planck measurement \citep{2020A&A...641A...6P} in our analysis.

\item We use model-independent lower limits on the ionization fraction (at $z \lesssim 6$) obtained from dark pixel fraction in quasar spectra \citep{2015MNRAS.447..499M}. The $1\sigma$ limits on the ionization fractions are $\ge 0.94-0.05$ at $z=5.9$ and $\ge 0.96-0.05$ at $z=5.6$. We utilize these limits to construct the likelihood.

\item We assume the reionization to end at $z \geq 5.3$. This limit is motivated by the recent observations of the Ly$\alpha$ optical depth from distant quasars \citep{2018MNRAS.479.1055B,2017ApJ...840...24E,2018ApJ...864...53E,2021arXiv210913170C} and theoretical models \citep{2019MNRAS.485L..24K,2021MNRAS.501.5782C,2020MNRAS.494.3080N,2020MNRAS.497..906K,2021arXiv210803699B}.

\item  We also utilize the galaxy UVLF data at $z = 6$ and $7$ obtained from optical studies \citep{2015ApJ...803...34B,2017ApJ...843..129B}.  

\item Lastly, we incorporate the temperature estimate at low density IGM as an additional probe of reionization history. There exist recent estimates of the $T-\Delta$ power-law relation, parametrized by $T_0$ and $\gamma$,  at $z = 5.4, 5.6$ and $5.8$ \citep{2020MNRAS.494.5091G} which serve our purpose. The estimated values for $T_0$ are $11000\pm1600~K$, $10500\pm2100~K$, and $12000\pm2200~K$ while $\gamma$ values are $1.20\pm 0.18$, $1.28\pm0.19$, and $1.04\pm0.22$ respectively for the above mentioned redshifts. These are estimated using the spike statistics of the Ly$\alpha$ transmitted flux.

\end{enumerate}
There are several other indirect constraints on the ionization fraction, e.g., those obtained from clustering of Ly$\alpha$ emitters \citep[e.g.][]{2022ApJ...940..102P}, Ly-break galaxies \citep[e.g.][]{2020ApJ...892..109N}, Ly$\alpha$ damping wings \citep[e.g.][]{2022MNRAS.512.5390G} and effective optical depth of the Ly$\alpha$/Ly$\beta$ forests \citep[e.g.][]{2021ApJ...923..223Z}. However, these constraints are highly model dependent. For example, the constraints obtained from the Ly$\alpha$ emitters depend on the patchiness of the ionized regions, apart from intrinsic properties of the Ly$\alpha$ emitting galaxies. Since the algorithm for generating the ionization maps in our semi-numerical code is different from the others, a self-consistent analysis would require computing the Ly$\alpha$ opacity arising from patchy neutral islands using our model, which is beyond the scope of this work. Similarly, the damping wing studies too require accurate modelling of the Ly$\alpha$ opacities in the quasar proximity zones, which in turn require extensions to our model beyond what can be done in this work. Hence, we do not use these observations in the likelihood for this work.
\subsection{Parameter constraints using the full MCMC}
\label{sec:analysis_mcmc}

For exploring the parameter space, we need to compute the likelihood for any given parameter vector $\mathbf{a}$. We use the standard  multidimensional gaussian likelihood ($ \mathcal{L}$) defined as 
\be
\label{eq:like_eq}
-2\ln \mathcal{L}(\mathbf{a}) = \chi^2(\mathbf{a}) = \sum_{i}\left[\frac{\mathcal{D}_i-\mathcal{M}_i(\mathbf{a})}{\sigma_i}\right]^2
\ee
where $\mathcal{D}_i$ are the measured values of the data points, $\mathcal{M}_i(\mathbf{a})$ are the model estimates for the parameters $\mathbf{a}$ and $\sigma_i$ are the observational error bars on the data. The summation index $i$ runs over all data points used in the analysis. Ideally, one should use the full error covariance matrix while computing the likelihood, however, it is unlikely to affect the analysis much as the used observational data errors are expected to be mostly uncorrelated. Hence, this form of the likelihood serves the purpose of this work in the absence of full covariance information from the observational estimates. For asymmetric errorbars, we use the upper uncertainty if the model estimate is above the data point and similarly use the lower uncertainty if the model estimate is below the data point.

The MCMC method employed in this work uses the Metropolis-Hastings algorithm \citep{1953JChPh..21.1087M} for sampling the parameter space and computing the posterior distribution. We utilize the publicly available package \textsc{cobaya} \citep{2021JCAP...05..057T}\footnote{\url{https://cobaya.readthedocs.io/en/latest/}} to perform the MCMC analysis.  The samples are drawn using 20 parallel chains \citep{2002PhRvD..66j3511L,2013PhRvD..87j3529L}. We assume the chains to be to converged when the Gelman-Rubin $R - 1$ value \citep{1992StaSc...7..457G} becomes less than a threshold $0.01$.  The first $30\%$ steps are removed from the chains as `burn-in'.

For this work, we use two sets of simulations, (i) one with a grid size $\Delta x = 64 h^{-1}$~cMpc having $4^3$ grid cells, which we call the `coarse resolution' simulation, and (ii) another with $\Delta x = 8 h^{-1}$~cMpc having $32^3$ grid cells, called the `high resolution' simulation. We were able to complete the full MCMC runs for these two cases. For these, we set wide priors for all the parameters:
\begin{itemize}
\item $\log \zeta_0 \in [0,10]$,
\item $\alpha \in [-20,20]$,
\item $\log T_{\mathrm{re}}\in [2,4.7]$,
\item $\log f_{\mathrm{esc}}^0 \in [-5,0]$ and
\item $\beta \in [-1,0]$.
\end{itemize}
A narrower prior range may discard some of the extreme reionization scenarios which can be allowed otherwise. Hence, we choose these wide prior ranges. Also, the wide priors can provide a robust analysis even if the data are modified in the future. A slightly narrower prior range is unlikely to create any impact on our results. Each of the two MCMC runs required a total of $\sim 10^5$ evaluations of the likelihood (hence that many calls to the semi-numerical simulation). It took only about 2-3 hours for the coarse resolution case to converge, while the time was much longer, about 6-7 days, for the high resolution. The parameter constraints for the two cases are shown in \fig{fig:mcmc_real_dat} (coarse resolution) and \fig{fig:mcmc_coarse_fine_jump} (high resolution) by red lines and contours. The implications of the parameter constraints, the correlations between the different parameters and other features of the posterior distributions have already been discussed in \citet{2022MNRAS.515..617M}, so we do not repeat them here.

The importance of these full MCMC runs is that the results obtained using the emulator will be compared against these. We emphasize here that producing the results for the coarse resolution in \fig{fig:mcmc_real_dat} took negligible computing time compared to those for the high resolution in \fig{fig:mcmc_coarse_fine_jump}, which will become important for the emulator to be discussed in the subsequent section.

\section{Building the likelihood emulator}
\label{sec:build_interp}
This section is devoted to building the likelihood emulator for our reionization model using GPR.

\subsection{A brief on Gaussian Process Regression (GPR) training}

Let us begin by providing a description
of the Gaussian Process (GP) Regression technique, which is used for training the likelihood emulator or interpolator. GP Regression (or GPR) is a non-parametric, Bayesian method that utilizes a set of random variables with a joint Gaussian distribution to predict the values of continuous quantities. This approach provides a probability distribution of all plausible functions that fit a given data set, without being limited by the choice of functional form. In general, the joint distribution is characterized by its covariance function or kernel in function space \citep{rasmussen}. The free parameters that describe the GP kernel form a hyper-parameter vector $\mathbf{h}$ that needs to be estimated from the characteristics of the training data. The dimension of the hyper-parameter space may vary depending on the choice of kernels and problem requirements.

In this analysis, we use the anisotropic \texttt{Matern} kernel (of order, $\nu=3/2$), in which each parameter direction has its own scaling hyperparameter. The goal is to find the interpolated $\chi^2(\mathbf{a})$ for any given set of parameter vector $\mathbf{a}$ utilizing the input distribution of a sample of $\{\chi^2,\mathbf{a}\}$ values (the training set). To build the trained interpolator, we use the methodology provided by the publicly available code \textsc{picasa} \citep{2022arXiv220507906P}. One of the main aspects of this code is to train a GP for the likelihood using a sparse sample of parameter points. This approach avoids the expensive computation of the cost function or likelihood at arbitrary parameter values. Once the training is successful, the interpolator is used to explore the parameter space via standard MCMC, but with enhanced speed.

We use only the functionality of the \textsc{picasa} framework relevant for GP training, as described next.
Given the input training set $\{\chi^2, \mathbf{a}\}$, the hyper-parameter vector $\mathbf{h}$ is optimised iteratively. At each iteration, starting with a small subsample of the training set, the log-marginal likelihood of the GP is maximised in hyper-parameter space using Algorithm 2.1 of \citet{rasmussen} as implemented in Scikit-Learn \citep{scikit-learn}.\footnote{\url{https://scikit-learn.org/}} In principle, this can be accomplished using any robust multi-dimensional minimisation algorithm; for convenience, we use the Anisotropic Simulated Annealing (ASA) algorithm  \citep{2022arXiv220507906P} packaged with \textsc{picasa}. This optimisation of $\mathbf{h}$
is followed by cross-validation of the corresponding GP
as a check of its accuracy, by using this GP to predict the values of $\chi^2$ for the part of the training sample not used at this iteration. If the cross-validation is not successful, the code proceeds to the next iteration using a slightly larger fraction of the training sample as input and exploring a slightly larger region of hyper-parameter space around the current optimum $\mathbf{h}$.
The cross-validation threshold can be tuned by a parameter \texttt{cv\_thresh} according to the required accuracy. If the 1 and 99 percentiles of the relative difference between the interpolated and actual $\chi^2$ is below the chosen threshold, the GP is assumed to be fully trained. This parameter also provides an estimate of the emulation errors for different cases.
 The values of the hyper-parameter components of the GP kernel, the final training size and training subsample, and cross-validation percentiles after the full training are stored as output. The trained GP, or likelihood emulator,
is further fed into the final MCMC run using the \textsc{cobaya} framework, which provides the desired posterior distribution of the free parameters. For further details of the GP training, we refer the reader to \citet{2022arXiv220507906P}.

The uniqueness of our method lies in training the emulator at the $\chi^2$ level, unlike the common approach of building up emulators using observables. 
We made this choice for the reasons discussed in the Introduction. Of course, this must be balanced with the fact that a new data set would require the GP to be re-trained to reflect the updated likelihood. Below we will assess the extent to which this can be done without running new simulations.

It is evident from the above discussion that the number of times one needs to run the simulation is equal to the size of the training set. The efficiency of the emulator thus would be determined by how small is the size of the training set compared to the number of steps required to run the MCMC. As we will see later, we can obtain speed-ups by factors $\sim 30$ using this method.

\begin{figure*}
    \centering
    \includegraphics[width=\textwidth]{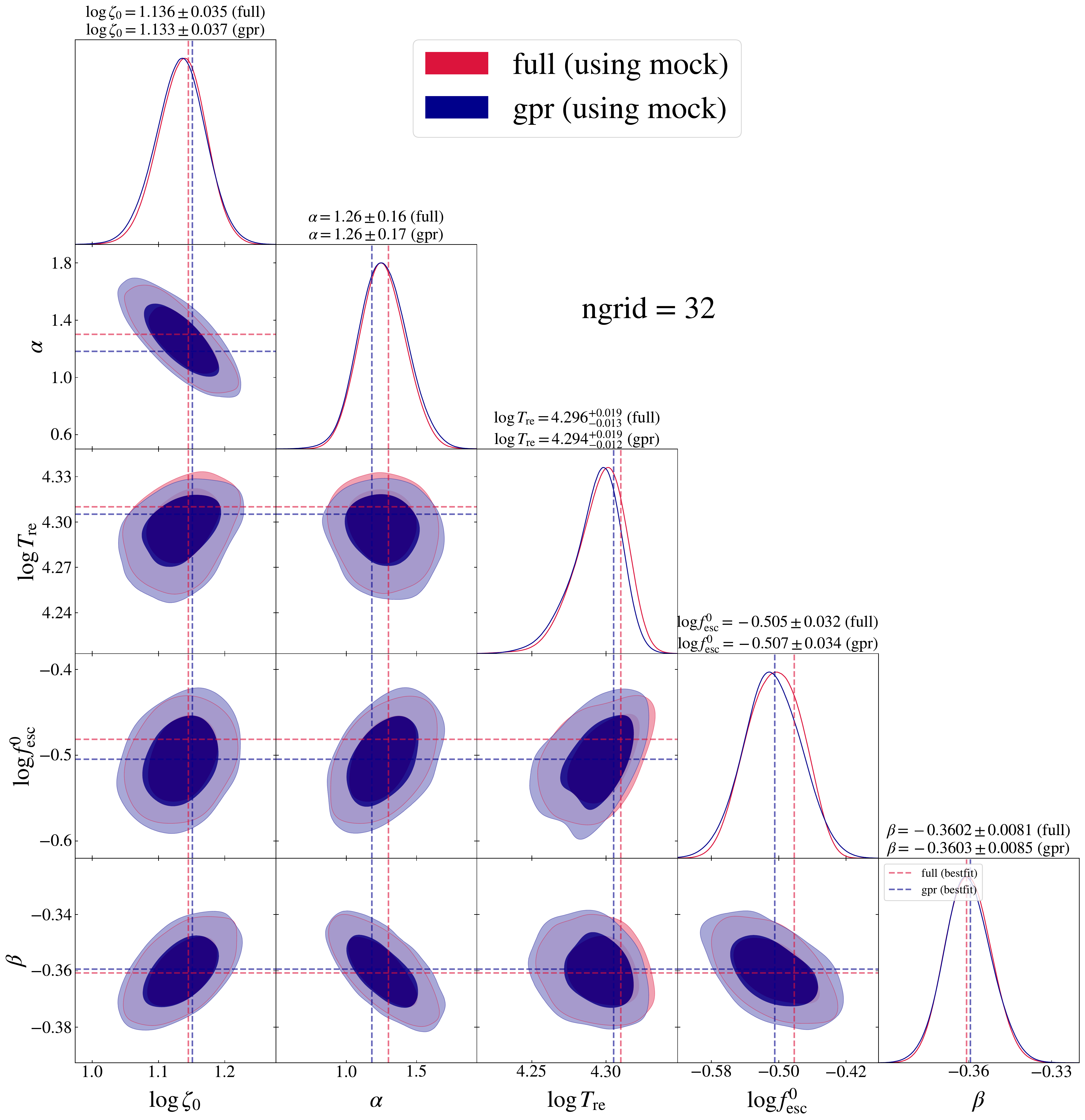}
    \caption{Comparison between full MCMC runs (ngrid=32) and GPR-trained MCMC using a standard smaller set of sample got from earlier analysis with real observational data. The data is generated from a \textbf{ fiducial mocks with shrunken errorbars instead of real available observations}.
    The formatting is identical to Fig.~\ref{fig:mcmc_real_dat}.
    }
    \label{fig:mcmc_mock_dat}
\end{figure*}

\subsection{The training set}

As discussed above, an essential aspect of training the interpolator with a limited number of samples is to select an effective training set. However, since we use broad priors on the free parameters, it is quite challenging to sample the entire parameter space keeping the sample size to within, say, a few thousand. As a result, methods like Latin hypercube sampling are not suitable for our case. 
Indeed, we found that the ASA algorithm, which is based on Latin hypercube sampling and is implemented in \textsc{picasa} by default, did not give us convergent results for our primary 5-dimensional model. Training convergence is much easier if the drawn sample has some information about the posterior distribution. Physically, it happens because the MCMC chain always spends more time near the high probability peak than the tail. This enhances the chance of error accumulation near the probability peaks while doing interpolation. Training with a relatively larger sample in those regions helps in keeping the errors within control. Ideally, to ensure that the training set is concentrated on regions of high probability and also that the samples trace the parameter degeneracies, one would need to know the posterior distribution of the parameters, which in turn would require an MCMC output or something equivalent to be available. 

We address this challenge by realising two points: firstly, the large-scale properties of our model are convergent with respect to the resolution and, secondly, the full MCMC runs with coarse resolution simulations require negligible computational cost. \emph{Hence, we use the chains from the coarse MCMC to construct the input training set.} We find that the training converges using a training set of size $\sim 3000-4000$. Therefore we need to run the high resolution simulation at most a few thousand times for the training. In fact, for most cases of our interest, the GP training itself is typically performed using only a small fraction of the training sample, making the GP evaluation of the likelihood extremely fast as compared to the full simulation. Once trained, the MCMC can proceed without any further simulations. This approach results in parameter constraints with only a few thousand calls to the simulation, as opposed to a total of around $10^5$ for a full MCMC run.

\section{Parameter constraints using the likelihood emulator}

\subsection{Comparison using available data}
\label{sec:real_dat}

We now present the results by comparing the posteriors obtained using our 
likelihood
emulator with those obtained from the full MCMC run.

\subsubsection{Setting up the emulator using a coarse resolution simulation}

To validate our method, we first test it on real observational data using a coarse resolution simulation with only $4^3$ grid cells. We use the full MCMC run of the coarse resolution simulation to construct the training set, train the emulator using the same coarse resolution simulation, and then compare the resulting parameter constraints with the full MCMC. This exercise may seem reduntant, but it serves the purpose of testing different aspects of the training method efficiently, given that the coarse resolution runs have negligible computational cost. Additionally, it enables us to identify the optimal number and distribution of input samples required for the training to converge, which will be used for training the high-resolution simulation.

In \fig{fig:mcmc_real_dat}, we show the comparison of the parameter distributions between original full MCMC run (red) and the fast MCMC run using the trained interpolator (blue). The full run takes a total of  $\sim 10^5$ likelihood evaluations to converge. We sample a subset of these to construct the training set, it turns out that the training converges using only 3175 evaluations with a moderate \texttt{cv\_thresh} of $0.04$. This reduction in the total number of function evaluations provides an efficiency which is more than one order of magnitude higher than the full runs. 

It is worth highlighting here that the priors for the GPR trained MCMC run need to be set up in such way that the interpolator remains confined to parameter space regions where the training was performed. As the small training set we use does not necessarily cover the entire parameter space uniformly, wide priors can lead to a non-convergent MCMC run post GP training. So, we choose the priors to be $\sim 3\sigma$ wide on either side of the mean of full MCMC run for each parameter, which covers the bulk of the high probability region. 

From \fig{fig:mcmc_real_dat}, it is clear that the best-fit values for the two cases are in good agreement with each other. The trained MCMC is also able to reproduce the two-dimensional joint probability distributions (68\% and  95\% confidence limits) along with the marginalised posterior distributions for individual parameters. 

\subsubsection{Emulating high resolution simulations}

This fast exploration of the parameter space can be particularly beneficial for the high resolution simulation runs, where the computation required for the likelihood evaluation can be considerably expensive. Although these high resolution runs are not that essential for the observables used in this work, it is important to set up the emulator for these expensive simulations for applications in the future. For instance, high resolution runs will be imperative to access the relatively high-$k$ (i.e., small-scale) modes of the 21~cm power spectra data expected in the next generation of telescopes.

Thus, our next goal is to evaluate the prospect of this novel method for higher resolutions (involving a higher number of grid cells, $32^3$). We begin by choosing the \emph{same} 3175 parameter locations as was used to train the GP using the coarse resolution simulation in the previous section, and compute the $\chi^2$ using the high resolution simulation only for these parameter values. This set of 3175 $\{\chi^2, \mathbf{a}\}$ serve as the training set.

We fix the \texttt{cv\_thresh} at $0.06$ which is slightly less stringent to ensure the convergence of training.\footnote{The value of \texttt{cv\_thresh} was fixed using a trial-and-error approach. Note that the computational cost for trying out different \texttt{cv\_thresh} values is negligible as they do not require any additional calls to the simulation.} The GP training converges with a training size of only a few hundreds. The trained interpolator is then used to run a fast MCMC. We emphasize that the fast MCMC in this case required 
only
3175 evaluations of the high resolution simulation, more than an order of magnitude smaller than what is required for a full MCMC run. 

In \fig{fig:mcmc_coarse_fine_jump}, we show the comparison of the posteriors between original full MCMC at high resolution (red) and the MCMC with trained interpolator (blue). 
The priors for the trained MCMC were the same as set for the coarse-resolution analysis.
As evident from the figure, the parameter posteriors obtained using both methods match each other exceptionally well. The best-fit values, shown by dashed lines, show some discrepancy for the two parameters $\alpha$ and $\log f_\mathrm{esc}^0$, however, the mismatch is well below the uncertainties in the parameters.

This result is encouraging as it indicates that the full MCMC at high resolution can potentially be bypassed by a more efficient exploration of the parameter space using the trained interpolator, even for a high-dimensional parameter space.

\subsection{Forecasting using mock data}
\label{sec:forecast}
\begin{table*}
\caption{Fiducial parameter values used to generate the mock data from near best-fit and shifted from best-fit scenario, along with priors chosen to run the post GPR training MCMC.}
\begin{threeparttable}
\begin{tabular}{cccc}
\hline

Parameters & Prior & Mock near best-fit (fiducial) & Mock shifted from best-fit (fiducial)\\
\hline
$\log\zeta_0$     & [$0.7,1.3$]     & $1.122$ & $0.96$  \\ \\
$\alpha$ & [$0.5,3.2$]  & $1.231$ & $1.963$\\ \\
$\log T_{\mathrm{re}}$     & [$4.11, 4.36$]      & $4.289$ & $4.198$ \\ \\
$\log f_{\mathrm{esc}}^0$   & [$-0.62,-0.2$]       & $-0.523$ & $-0.42$     \\ \\
$\beta$     &  [$-0.5,-0.32$]         & $-0.361$ & $-0.418$   \\ \\
\hline
\end{tabular}
\end{threeparttable}
\label{tab:param_cons}
\end{table*}

Up to this point, we have checked the performance of our emulator in constraining parameters using currently available data. However, since we train the emulator using $\chi^2$ values obtained from a specific data set, it raises the question of whether the 
training 
samples 
will remain useful if the data set changes. In the near future, we can expect high-quality observations to be added to the existing ones, such as improved measurements of the UVLF using the JWST \citep{2023MNRAS.518.6011D,2023arXiv230414469M}, and tighter constraints on $\tau_e$ from next-generation CMB experiments, such as LiteBIRD \citep{2022arXiv220202773L}.

To address this issue, we 
assume that with the availability of higher quality data, the errors associated with all the observables used in this work will 
decrease. For simplicity, here we assume a reduction by a constant factor 2 in each error bar.
We then generate mock data for this scenario and repeat the analysis, using the same training set as in the previous section.

\subsubsection{Mock near best-fit}

While generating the mock data, we must assume an underlying model. In this section, we assume that the model remains the same as the best-fit model obtained in the previous analysis and compute the observables. We introduce noise in the observables by adding a gaussian random number with zero mean and standard deviation equal to the updated errorbars. For asymmetric errorbars, we shift the values by a gaussian random number with a mean of zero and standard deviation equal to 10\% of the mock value. This randomization accounts for the inherent uncertainties associated with instrumental measurement. 

Next, we calculate updated $\chi^2$ values using this mock data at the positions where the earlier training parameters are stored. This corresponds to only a few thousand (specifically, 3175) runs of the high resolution simulation. In fact, these calls to the simulation can also be by-passed by simply storing all the observables at the parameter locations of the training set.
We take a subset of these samples by removing the parameters which lead to a large $\chi^2$ value ($\ge 50$) and use these as our new training set. These very large $\chi^2$ values are not suitable for training the GPR because the emulator finds it difficult to interpolate between such high values. Thus, we remove these parameter vectors to ensure that the training procedure provides convergence. We perform the training with the remaining $\sim 1000$ sample points, which are sufficient to provide the converged GPR. In fact, the training converges with only 449 sample evaluations for a \texttt{cv\_thresh} of 0.18. The priors are kept the same as in the previous section.

In parallel, we also run the standard full MCMC with this mock data, which will act as a benchmark. In \fig{fig:mcmc_mock_dat}, we compare the expected posteriors using mock data between the full MCMC run (red) and trained MCMC run (blue). The first point to note is that the constraints on the parameters have significantly tightened compared to the analysis done using the presently available data, a direct consequence of the reduced errorbars. It can also be seen that the best-fit values, along with the one-dimensional posterior distributions, show excellent agreement with each other. The joint probability contours from the trained MCMC (68\% and 95\%) are also similar to the standard MCMC. 

This analysis demonstrates that the emulator, developed with the present available data in mind, can perform equally well when the observational errorbars shrink in the future, as long as the best-fit value remains close to its present value.

\subsubsection{Mock shifted from best-fit}

As the performance of the emulator proved to be satisfactory when the mock data was generated using the current best-fit model, we next perform a similar analysis assuming that the best-fit model in the future may deviate from its current value. To achieve this, we generate mock data using a set of parameters which are around $1\sigma$ away from the best-fit values obtained using the presently available data. Following similar method as stated in the earlier subsection, we reduce the errorbars by a factor of 2 and add gaussian random noise on top of the mean values.

\begin{figure*}
    \centering
    \includegraphics[width=\textwidth]{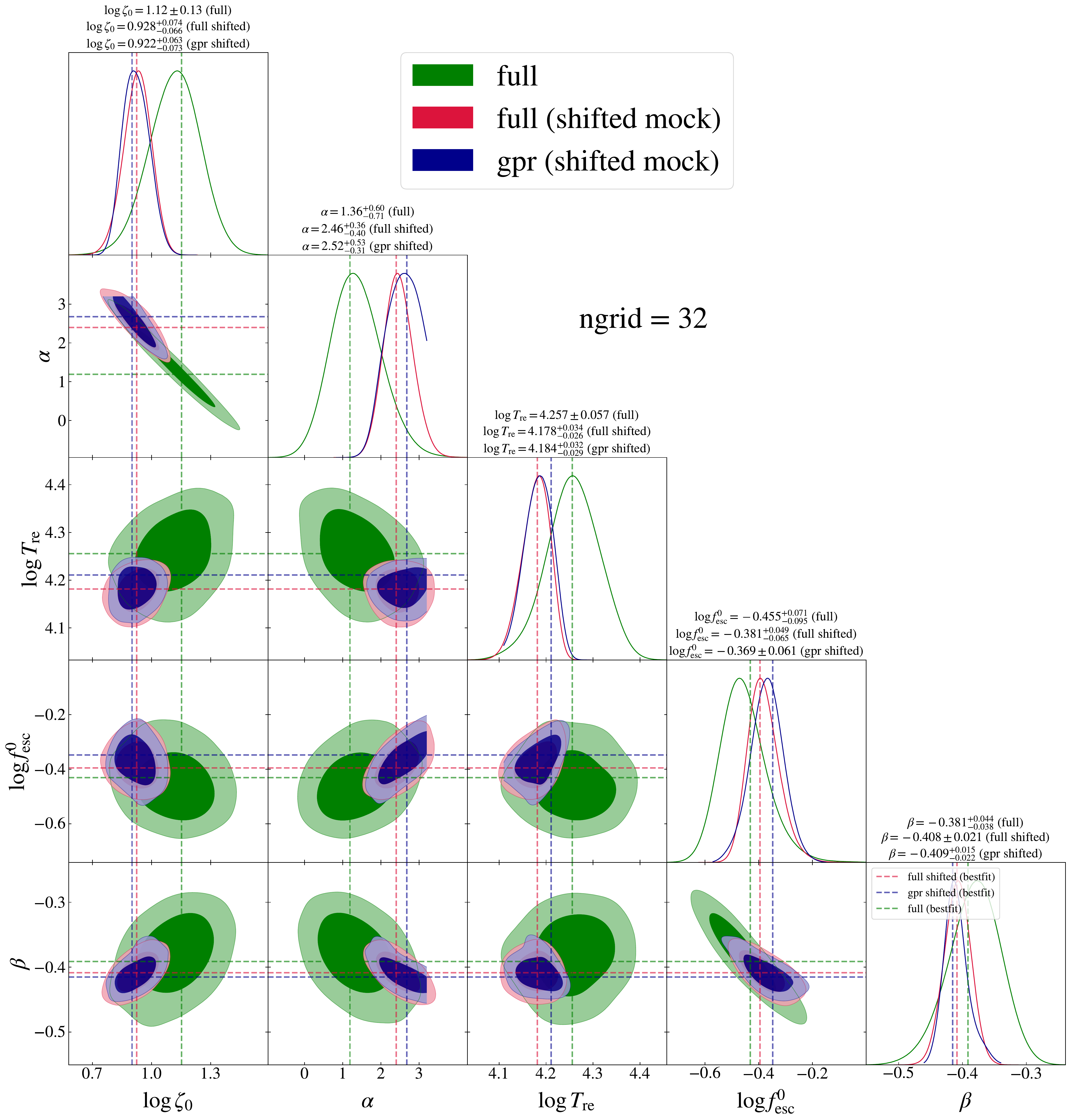}
    \caption{Comparison between full MCMC (\textit{red}) runs (ngrid=32) and GPR-trained MCMC (\textit{blue}) using a standard smaller set of sample got from earlier analysis with real observational data. The data is generated from a \textbf{ fiducial mocks (1$\sigma$ shifted from best-fit)  with shrunken errorbars instead of real available observations}.  
    The formatting is identical to Fig.~\ref{fig:mcmc_real_dat}. The \textit{green} contours correspond to the analysis  with real observational data.} 
    \label{fig:mcmc_mock_dat_sig}
\end{figure*}

In \fig{fig:mcmc_mock_dat_sig}, we present the posterior distributions of the full MCMC (red) and the trained MCMC (blue) using the mock data. For comparison, we also show the original MCMC using the presently available data (green; same as the \textbf{full} posterior, shown in red, in \fig{fig:mcmc_coarse_fine_jump}). The new mock data was used to compute $\chi^2$ values at the same 3175 parameter locations, and we removed the parameters with $\chi^2 \ge 50$ to ensure convergence during training. We also use a similar \texttt{cv\_thresh} of 0.15 for the training. The training process requires around $\sim 300$ function evaluations. We again keep the priors same as discussed in section \ref{sec:real_dat}.

The results demonstrate that both the full MCMC and trained MCMC provide a good fit for most of the parameters. However, due to the scarcity of available training data points, the constraint on $\alpha$ exhibits an abrupt cut at higher values. To elucidate this behavior, we can compare the posterior distributions obtained using the full MCMC for the presently available data (green) and the mock data (red). We observe that the contours for the mock data stretch beyond those for the presently available data, especially for high values of $\alpha$. Since the training samples were drawn solely from the green probability distributions, it is clear that the emulator did not receive any information on the $\chi^2$ values for regions beyond these green contours. Consequently, it is not surprising that the trained GP fails to provide the correct parameter posteriors for high $\alpha$ values. However, we must emphasize that the emulator failing for the parameter $\alpha$ is merely a result of where we fixed the best-fit value for the mocks; otherwise, there is nothing peculiar about $\alpha$.

This analysis highlights a limitation of our method: the emulator may fail to provide accurate results when the true posteriors extend beyond the distribution used for constructing the training sample. In such cases, it is necessary to retrain the emulator (i.e., perform the entire analysis described in section~\ref{sec:real_dat}) using the updated data set to ensure its reliability and accuracy. We are currently exploring techniques to ensure maximum overlap between the old and new training sets in such cases, so as to minimise the number of new simulation calls.

\section{Summary \& Conclusions}
\label{sec:conc}

In this work, we explore the potential of a new method for efficiently exploring parameter space using the semi-numerical photon conserving model, \texttt{SCRIPT} \citep{2022MNRAS.515..617M}. The method is based on the Gaussian Process (GP) Regression technique  to avoid the computationally expensive full MCMC exploration, along the lines discussed by \citet{2022arXiv220507906P}. The idea is to train 
a GP to emulate the likelihood for a given data set with a relatively small set of parameter samples, which in turn are obtained from a low-cost fast MCMC, and then use the trained likelihood emulator for the exploration of the parameter space. Our motivation for this approach arises from the need to explore the reionization parameter space using high-resolution semi-numerical simulations, which are computationally expensive. For instance, upcoming interferometers such as SKA-Low and HERA will provide a wide range of 21~cm power spectra Fourier modes, and accessing these modes would require running high-resolution simulations with a number of grids of $128^3$, which is impractical for full MCMC runs. 

Let us summarize the main findings of our study:

\begin{itemize}

    \item We started by studying a coarse resolution simulation ($4^3$ grids) with available observational constraints and found that the trained emulator can provide parameter contours comparable with the full MCMC runs. The parameter exploration with this fast method provides about an order of magnitude speedup compared to the full run.

    \item Next, we performed a feasibility analysis of our method for simulation with a resolution eight times higher ($32^3$ grids). We used the same training parameter set as in the coarse resolution analysis and computed the modified $\chi^2$ at the higher resolution. The analysis shows that the trained emulator using a small set of samples can again perform as well as the full MCMC exploration. This opens up a new way for efficient parameter space exploration using a training parameter set drawn from fast, coarse-resolution runs and then modifying the $\chi^2$ values corresponding to that smaller set for training at higher resolution.

    \item Lastly, we conducted a parameter forecast study with mock data assuming futuristic observational constraints. This provided an idea about the applicability of our technique in the near future when more stringent constraints will be available. We found that the trained emulator can reproduce the confidence intervals and related contours comparable to the full MCMC when the mock data is generated with parameter values near the previous best-fit of the parent MCMC run with available data. In case the location of the best-fit significantly deviated from the present value, the emulator needs to be retrained.

 \end{itemize}

Our method of constructing a training set based on the output of an MCMC using coarse resolution simulations has similarities to the GP-based approach of \citet{2017ApJ...848...23K} where they converge towards the region around the best-fit model by running several MCMCs iteratively. It would be interesting to compare the efficiency of these two methods in terms of the number of simulation calls required to obtain a reliable training set, particularly for a similar number of free parameters. Additionally, it would be worthwhile to extend our calculations for artificial neural network based emulator methods such as those used by \citet{2018MNRAS.475.1213S} who were able to achieve an emulator using only $\sim 100$ samples of the 21~cm power spectra.

To summarize, our new method of training a likelihood emulator can greatly improve the efficiency of reionization parameter space exploration when the likelihood computation is costly. In this paper, we have demonstrated the potential of this method and discussed its possible advantages and limitations. Moving forward, we plan to utilize this technique with 21 cm data (either mock or real) to fully test its potential.

\section*{Acknowledgements}

BM and TRC authors acknowledge support of the Department of Atomic Energy, Government of India, under project no. 12-R\&D-TFR-5.02-0700. The research of AP is supported by the Associateship Scheme of ICTP, Trieste.

\section*{Data Availability}

A basic version of the code, which does not include the effects of recombinations and feedback on ionization maps, used in the paper is publicly available at \url{https://bitbucket.org/rctirthankar/script}. The code \textsc{picasa} is publicly available at \url{https://bitbucket.org/aparanjape/picasa/}. The data obtained from the extensions of the code and presented in this article will be shared on reasonable request to the corresponding author (BM).



\bibliographystyle{mnras}
\bibliography{script_picasa_rev2} 







\bsp	
\label{lastpage}
\end{document}